\documentclass[preprint]{aastex}

\begin{document}

\slugcomment{The Astrophysical Journal, accepted 2001 November 19.}

\title{Detection of an Extrasolar Planet Atmosphere\altaffilmark{1}}
\author{David Charbonneau\altaffilmark{2,3}, Timothy M. Brown\altaffilmark{4},
Robert W. Noyes\altaffilmark{3}, \and Ronald L. Gilliland\altaffilmark{5}}
\altaffiltext{1}{Based on observations with the NASA/ESA
{\it Hubble Space Telescope}, obtained at the Space
Telescope Science Institute, which is operated by the
Association of Universities for Research in
Astronomy, Inc. under NASA contract No. NAS5-26555.}
\altaffiltext{2}{California Institute of Technology,
105-24 (Astronomy), 1200 E. California Blvd., Pasadena, CA 91125;
dc@astro.caltech.edu.}
\altaffiltext{3}{Harvard-Smithsonian Center for Astrophysics,
60 Garden St., Cambridge, MA 02138;
rnoyes@cfa.harvard.edu.}
\altaffiltext{4}{High Altitude Observatory/National Center for Atmospheric
Research, 3450 Mitchell Lane, Boulder, CO 80307; timbrown@hao.ucar.edu.  
The National Center for Atmospheric Research is sponsored
by the National Science Foundation.}
\altaffiltext{5}{Space Telescope Science Institute, 3700 San Martin Dr.,
Baltimore, MD 21218; gillil@stsci.edu.}

\begin{abstract}
We report high precision spectrophotometric observations
of four planetary transits of HD~209458, 
in the region of the sodium 
resonance doublet at 589.3~nm.  We find
that the photometric dimming during transit
in a bandpass centered on the sodium feature is deeper by 
$(2.32 \pm 0.57) \times 10^{-4}$ relative to simultaneous
observations of the transit in adjacent bands.  We interpret this
additional dimming as absorption from sodium in the planetary 
atmosphere, as recently predicted from several theoretical modeling
efforts.  Our model for a cloudless planetary 
atmosphere with a solar abundance of sodium in atomic form 
predicts more sodium absorption than we observe.
There are several possibilities that may account for this reduced 
amplitude, including reaction of atomic sodium
into molecular gases and/or condensates, 
photoionization of sodium by the stellar
flux, a low primordial abundance of sodium, or the presence 
of clouds high in the atmosphere. 
\end{abstract}


\section{Introduction}
Since the discovery of planetary transits in the light curve of 
HD~209458 \citep{cha00,hen00,maz00}, this star has been the subject
of intensive study:  Multicolor observations \citep{jha00, dee01}
have confirmed the expected color-dependence of the light curve
due to the stellar limb-darkening.  Radial-velocity monitoring during
transit \citep{que00, bun00} has yielded variations in excess of
the orbital motion, due to the occultation of the rotating stellar 
limb by the planet.  Very-high-precision photometry 
\citep[][hereafter B01]{bro01b}
permitted an improved estimate of the planetary and stellar
radii, orbital inclination, and stellar limb-darkening, as well as
a search for planetary satellites and circumplanetary rings.

Transiting extrasolar planets present a unique opportunity for us
to learn about the atmospheres of the objects.  Wavelength-dependent 
variations in the height at which the planet becomes
opaque to tangential rays will result in wavelength-dependent
changes in the ratio of spectra taken in and out of transit.
Several groups
\citep{sea00, bro01a, hub01} have pursued theoretical explorations
of this effect for a variety of model planetary atmospheres.  
These studies have demonstrated that 
clouds, varying temperature structure and chemical composition, 
and even atmospheric winds, all produce variations that would
be observed, should the requisite precision be achieved.
Based on these calculations, the expected variations could
be as large as 0.1\% relative to the stellar continuum.

A previous search by \citet{bun00} 
for variations between spectra of HD~209458
observed in and out of transit (using extant spectra gathered for 
radial velocity measurements) was sensitive to features
with a relative intensity greater than 1--2\% of the stellar continuum.
\citet{mou01} conducted a similar study, and achieved a detection
threshold of 1\%.  Although the precision achieved by
both these studies was insufficient to address any reasonable model
of the planetary atmosphere, \cite{mou01} placed limits on the models
of the planetary exosphere.  Searches for absorption due to
a planetary exosphere have also been conducted for 51~Peg 
during the times of inferior conjunction \citep{cou98, rau00}.

\section{Observations and Data Analysis \label{data_section}}
We obtained 684 spectra\footnote{These data are
publicly available at http://archive.stsci.edu.} 
of HD~209458 with the \emph{HST} STIS spectrograph, 
spanning the times of four planetary transits,
on UT 2000 April 25, April 28--29, May 5--6, and May 12--13.  
The primary science goal of this project was to improve the 
estimate of the planetary and stellar radii, orbital inclination, 
and stellar limb-darkening, as well as conduct a search for 
planetary satellites and circumplanetary rings; 
we presented these results in B01.  In
selecting the wavelength range for these observations,
the desire to maximize the number of detected photons
led us to consider regions near 600~nm, 
where the combination of the instrumental sensitivity
and the stellar flux would be optimized.

The secondary science goal was to pursue the prediction
by \citet{sea00} \citep[and, later,][]{bro01a, hub01}
of a strong spectroscopic feature at 589.3~nm due
to absorption from sodium in the planetary atmosphere.
Thus we chose to observe the wavelength region
$581.3 \le \lambda \le 638.2$~nm, with a medium
resolution of $R = \lambda / \Delta \lambda = 5540$,
corresponding to a resolution element of 0.11~nm.

The details of the data acquisition and analysis
are presented in B01, and we refer the reader to that publication.
In summary, the data reduction consisted of (1) recalibrating
the two-dimensional CCD images, (2) removing cosmic-ray
events, (3) extracting one-dimensional spectra,
(4) summing the detected counts over wavelength
to yield a photometric index, and (5) correcting
the resulting photometric time series for variations
that depend on the phase of the \emph{HST} orbit
and for variations between visits.  
The only difference in these procedures between
B01 and the current work is in step (4):
Previously, we summed the spectra either (a)
over the entire available wavelength range, or 
(b) over the blue and red halves of the available range.  
In the present paper, 
we restrict greatly the wavelength span over
which we perform the integration.

As described in B01, observations of the first transit
(UT 2000 April 25) were partially compromised by a database
error in the location of the position of the spectrum
on the detector.  The result was that the spectrum
was not entirely contained within the CCD subarray.
In the subsequent data analysis, we ignored these
data.  Furthermore, as described in B01, the first orbit (of 5)
for each visit shows photometric variability in excess of that achieved 
for the remaining orbits.  These variations are presumably
due to the spacecraft; we omit these data as well.
This leaves 417 spectra of the 684 acquired.

Since we do not know the precise width of the feature we
seek, we select three bands of varying width, each centered
on the sodium feature.
We refer to these bands as ``narrow'' ($n$), ``medium'' ($m$), 
and ``wide'' ($w$).  The $n$ band is the smallest 
wavelength range that still encompasses the stellar sodium
lines; the $w$ band is the widest wavelength range that
permits an adjacent calibrating band to the blue.
The $m$ band is roughly intermediate (by ratio)
between these two extremes; it is $\sim 1/3$ the range of the
$w$ band, and $\sim 3$ times the range of the $n$ band.
We further define, for each of these, a ``blue'' ($b$) 
and a ``red'' ($r$) band,
which bracket the ``center'' ($c$) band.    The names and 
ranges of these nine bands are given in Table~1, and displayed
in Figure~1.  For each of these 9 bands, 
we produce a photometric time series by the procedure
described above. The photometric index at a 
time $t$ is then identified by a letter
indicating the width, and a subscript indicating the position (e.g.
``$w_r(t)$'' indicates the photometric index in the wide band, 
red side, at time $t$).  Each of these is a normalized 
time series, with a value of unity when averaged over the out-of-transit
observations; the minimum values near the transit centers 
are approximately 0.984.

We denote the time of the center of the photometric transit by $T_{c}$.  
In what follows, we consider the in-transit observations,
(i.e. those that occur between second and third contacts,
$|t-T_c| < 66.111$~min), which we denote 
by $t_{in}$, and the out-of-transit observations (those that occur before 
first contact, or after fourth contact, $|t-T_c| > 92.125$~min), 
which we denote by $t_{out}$.  There are 171 in-transit observations,
and 207 out-of-transit observations.  We ignore the
small fraction (7\%) of observations that occur during ingress
or egress.

\subsection{Stellar Limb-Darkening}

One potential source of color-dependent variation
in the transit shape is stellar limb-darkening 
\citep[see Figure~6 in B01; also][]{jha00, dee01}.
In order to investigate this possibility,
we produce (for each width) the difference of the red 
and blue bands:
\begin{equation}
\begin{array}{ccccc}
n_d(t) & = & n_b(t) & - & n_r(t)\phm{.}\\
m_d(t) & = & m_b(t) & - & m_r(t)\phm{.}\\
w_d(t) & = & w_b(t) & - & w_r(t).
\end{array}
\end{equation}
The observed standard deviations in these time series 
(measured over the out-of-transit observations) are 
$\sigma [n_{d}(t_{out})] = 3.3 \times 10^{-4}$, 
$\sigma [m_{d}(t_{out})] = 3.6 \times 10^{-4}$, 
and $\sigma [w_{d}(t_{out})] = 5.4 \times 10^{-4}$.
These values match the predictions based on photon noise.

In order to check for changes in the
transit depth due to stellar limb-darkening, 
we then calculate, for each of the time series above,
the difference in the mean of the relative flux,
between the in-transit observations and the
out-of-transit observations:
\begin{equation}
\begin{array}{ccccccc}
\Delta n_d & = & \overline{n_d(t_{in})} & - & \overline{n_d(t_{out})} & = & (-2.0 \pm 2.5) \times 10^{-5}\phm{.}\\
\Delta m_d & = & \overline{m_d(t_{in})} & - & \overline{m_d(t_{out})} & = & (+0.1 \pm 2.7) \times 10^{-5}\phm{.}\\
\Delta w_d & = & \overline{w_d(t_{in})} & - & \overline{w_d(t_{out})} & = & (-3.1 \pm 4.1) \times 10^{-5}.
\end{array}
\end{equation}
Overbars indicate averages over time, and each quoted error
is the estimated standard deviation of the mean.
All of these values are consistent with no variation.
Thus, we have no evidence for color-dependent limb-darkening
between these bands, which span about 15~nm.

We compared these results to predictions based on a 
theoretical model 
of the stellar surface brightness.  The model was
produced by R.~Kurucz (personal communication, 2000).  
We denote it by $S(\lambda,\mu)$, 
where $\mu$ is the cosine of the angle between the line of 
sight and the normal to the local stellar surface.  
The stellar parameters were taken to be the best-fit 
values as found by \citet{maz00}:
$T_{eff} = 6000$~K, $\log g = 4.25$, and $[ {\rm Fe} / {\rm H} ] = 0.00$.
The model is evaluated at a spectral resolution of 
$R = 2~000~000$ (greatly in excess of the
STIS resolution, $R = 5540$), and at 17~values
of $\mu \in [0,1]$

The stellar model allows us to produce a theoretical
transit curve over a chosen bandpass, which includes
the effects of stellar limb-darkening. 
We do not force the limb-darkening to fit a parameterized model.  
Rather, we interpolate between the values of $\mu$ on 
which the model of $S(\lambda,\mu)$ is initially calculated.  
For each $\lambda$, we produce the theoretical transit curve, 
$T(\lambda, t)$, for all $t$ at which we have data.
The curve is calculated by integrating over the unocculted portion of
the stellar disk as described in \citet{cha00} and
\citet{sac99}.  We assume the best-fit 
values from B01 for the stellar radius ($R_{s}=1.146~R_{\Sun}$), 
planetary radius ($R_{p}=1.347~R_{\rm Jup}$, 
where $R_{\rm Jup}=71492$~km
is the equatorial radius of Jupiter at a pressure of 
1~bar; \citet{cox00}), orbital period 
($P=3.52474$~days), orbital inclination ($i=86{\fdg}6$) 
and semi-major axis ($a=0.0468$~AU), 
assuming a value for the stellar mass of 
$M_{s} = 1.1~M_{\Sun}$ \citep{maz00}.

We note that the STIS data have an effective continuum
that is tilted relative to the theoretical values,
due to the wavelength dependent sensitivity of the instrument.
We wish to emulate the data, where each $\lambda$
is weighted by the observed number of photons. 
To do so, we fit a low-order polynomial (in $\lambda$) 
to the model disk-integrated spectrum, and a second 
low-order polynomial to the STIS data.  We subsequently 
divide the calculated $T(\lambda, t)$ for each
$t$ by the ratio of these polynomials.

Next, we integrate $T(\lambda, t)$ in $\lambda$ 
over each of the band passes in Table~1.
We subsequently difference the model light curve as calculated
for the blue and red bands (emulating what we did with
the data in eq.~[1]), and then evaluate the mean of the 
difference (as in eq.~[2]):
\begin{equation}
\begin{array}{cccc}
\Delta n_d & {\rm (theory)} & = & -0.44 \times 10^{-5}\phm{.}\\
\Delta m_d & {\rm (theory)} & = & -0.58 \times 10^{-5}\phm{.}\\
\Delta w_d & {\rm (theory)} & = & -0.85 \times 10^{-5}.
\end{array}
\end{equation}
These results can be compared directly with equation~(2).
Each of these is much smaller than
the observational precision we achieved, and thus 
consistent with our observational result of no significant
offset.

Furthermore, the results stated in equation~(2) demonstrate
that the data remain photon-noise limited at 
these high levels of precision:  We have taken
the mean of two large groups of data (roughly 200
observations apiece), and found that the
difference is consistent with photon-noise limited photometry, 
with a typical precision $3 \times 10^{-5}$.

\subsection{The Sodium Band}

In order to search for variations in the sodium band
relative to the adjacent bands, we produce (for each width)
the mean light curve of the blue and red bands, and difference 
this from the light curve for the center band:
\begin{equation}
\begin{array}{ccccccccc}
n_{Na}(t) & = & n_c(t) & - & [n_b(t) & + & n_r(t)] & / & 2\phm{.}\\
m_{Na}(t) & = & m_c(t) & - & [m_b(t) & + & m_r(t)] & / & 2\phm{.}\\
w_{Na}(t) & = & w_c(t) & - & [w_b(t) & + & w_r(t)] & / & 2.
\end{array}
\end{equation}
This linear combination removes the variations due to the color 
dependence of the limb-darkening of the stellar continuum.
As we found above (\S~2.1), this effect is very small 
(eq.~[3]).  We
consider the effect of the deviations in the
limb-darkening that occur in the cores of
stellar absorption features in \S~4.1, and
show that this effect is also negligible.
The three time series in equation~(4) are plotted as a function
of absolute value of the time from the center of transit
in Figure~2.
The observed standard deviations in these time series 
over the out-of-transit observations are 
$\sigma [n_{Na}(t_{out})] = 5.5 \times 10^{-4}$, 
$\sigma [m_{Na}(t_{out})] = 3.7 \times 10^{-4}$, 
and $\sigma [w_{Na}(t_{out})] = 3.5 \times 10^{-4}$.
These achieved values match the predictions of the 
photon-noise-limited precision.

In order to look for changes in the
transit depth in the sodium-band relative to the
adjacent bands, we calculate, for each of the time series above,
the difference in the mean as observed in and out of transit:
\begin{equation}
\begin{array}{ccccccc}
\Delta n_{Na} & = & \overline{n_{Na}(t_{in})} & - & \overline{n_{Na}(t_{out})} & = & (-23.2 \pm 5.7) \times 10^{-5}\phm{.}\\
\Delta m_{Na} & = & \overline{m_{Na}(t_{in})} & - & \overline{m_{Na}(t_{out})} & = & (-13.1 \pm 3.8) \times 10^{-5}\phm{.}\\
\Delta w_{Na} & = & \overline{w_{Na}(t_{in})} & - & \overline{w_{Na}(t_{out})} & = & (-{\phn}3.1 \pm 3.6) \times 10^{-5}.
\end{array}
\end{equation}
As before, the errors shown are the 1-$\sigma$ errors of the mean.
The results indicate that we have detected a deeper transit in the
sodium band for the narrow and medium bandwidths, with a significance 
of $4.1~\sigma$ and $3.4~\sigma$, respectively.  
We find no significant offset for the wide bandwidth.

Furthermore, we calculate the three quantities in equation~(5) for 
each of the three transits separately.  These results confirm 
the conclusion that there is a deeper signal in the narrow 
and medium bands.  The variations between visits for the same 
band are not significant given the precision of the data.

In Figure~3, we plot histograms of the
in-transit data and the out-of-transit data,
for the narrow band.  This plot shows that 
each of these sets of points appears to be drawn 
from a normal distribution with a standard deviation
as predicted by photon-counting arguments.
However, the in-transit points scatter
about a mean that is significantly offset from that
defined by the out-of-transit points.  This indicates that
we have not skewed these distributions by
our analysis procedures:  Rather, the in-transit
observations appear to represent a shifted version of
the out-of-transit observations.

We bin $n_{Na}(t)$ in time, and plot these results
in Figure~4.  These further illustrate that we have
observed a deeper transit in the sodium band.

We investigate the possibility that the observed
decrement is due to non-linearity in the STIS
CCD:  Since the observed value of $\Delta n_{Na}$ is 
$-2.32 \times 10^{-4}$ for a change in mean intensity 
inside to outside transit of $-1.6 \times 10^{-2}$, 
a non-linearity of $\sim 1$\% across this range would be required.
\citet{gil99} derive an upper limit that is an order of magnitude lower
than this.  We further test this effect directly
by selecting 18~strong stellar absorption features in our
observed wavelength range.  For each spectral feature,
we define a set of band passes with the same wavelength
range as those listed in Table~1, but now centered
on the spectral line.  We derive $\Delta n$ and 
$\Delta m$ for each of these as in equation~(5).
The results are shown in Figure~5.  
We find no correlation of either $\Delta n$
or $\Delta m$ with spectral line depth.
These tests also confirm our evaluation of the precision
we achieve in equation~(5):  We find that the values
of $\Delta n$ or $\Delta m$ are normally distributed, 
with a mean consistent with 0, and 
a standard deviation as we found above (eq.~[5]).

\section{Comparison with Theoretical Predictions}

As described above,
the quantity that we observe is the difference in the
transit depth in a band centered on the Na~D lines
to the average of two flanking bands, as a
function of time from the center of transit:
We denoted these by $n_{Na}(t)$,  $m_{Na}(t)$,
and  $w_{Na}(t)$, above.
We wish to compare these results to 
models of the planetary atmosphere.  Several
steps are required to transform the model predictions
into the same quantity as the observable, and
we describe these steps here.

We first produce 
several model calculations of the change in the effective
planetary radius as a function of wavelength, 
as prescribed by \citet{bro01a}.
For all the models, we 
use the best-fit values from B01 for $R_{p}$, $R_{s}$,
$i$, $a$, $M_{s}$ as given in \S~2.1.
We further specify the planetary mass ($M_{p} = 0.69~M_{\rm Jup}$)
and the stellar effective temperature ($T_{s} = 6000$~K),
both as given by \citet{maz00}.  We set the equatorial 
rotational velocity of
the planet to $v_{rot} = 2.0 {\rm \; km \, s^{-1}}$.
This is the value implied by the measured planetary
radius, under the assumption that the planet is tidally locked
(i.e. the rotational period is the same as the orbital period).
Each model includes the effect of photoionization 
as described in \citet{bro01a}, which reduces the core 
strength of the Na~D lines, but does not significantly 
change the wings.

The fiducial model has a cloud deck with cloud 
tops at a pressure of .0368 \mbox{( = 0.1 / e)}~bar, and solar metallicity
\citep[the measured value for HD~209458;][]{maz00}.
We also consider a number of variants to the fiducial
model, representing changes to either the metal abundance,
or the cloud height.  These variants are listed
in Table~2.  For each model (denoted by the index $m$),
we have the theoretical planetary radius
$R_{p}(\lambda, m)$, where $\lambda$ denotes
the wavelength. For each of these, we
renormalize $R_{p}(\lambda, m)$ so that the average
value of the planetary radius over the 
entire STIS bandpass is the best-fit value determined by 
B01 ($1.347~R_{\rm Jup}$).
These renormalized functions $R_{p}(\lambda, m)$
are shown in Figure~6.

We use the theoretical model of the stellar surface
brightness [$S(\lambda,\mu)$; described in \S~2.1] 
to include the effects of the wavelength-dependent 
limb-darkening of the star.  
This approach allows us to account for both the continuum 
limb-darkening, and the deviations that occur from this in the
cores of absorption lines; these can be significant
in the case of the sodium features.  
In general, the limb-darkening
is less pronounced in the cores of the sodium
lines than in the adjacent continuum, and thus 
for a planetary radius that is independent
of wavelength, the transit light curve near $T_{c}$ 
is deeper and more rounded as observed in the continuum 
than as observed in the core of an absorption line.

We proceed as described in \S~2.1, 
with the exception that $R_{p}(\lambda, m)$ 
is now a function of wavelength:
For each $\lambda$,
we produce the theoretical transit curve, $T(\lambda, t, m)$.
We bin these results in the wavelength bands
as described in \S~2, and
difference the values in the band centered on the sodium
lines from the average of the flanking bands,
as prescribed by equation~(4).
The result is a theoretical time series that closely
resembles the observed STIS time series, with the effect
of a wavelength-dependent planetary radius included.
These model time series are shown in Figure~7,
and may be compared directly with the data shown
in Figures~2 \& 4.

We find that most of the models listed in Table~1
produce differential transits that are significantly
deeper than our observational result.  Only models
n4 or c3, which represent, respectively, 
extreme values of high cloud height or depleted atomic sodium 
abundance, produce transits of approximately 
the correct depth.

\section{Discussion}

We have detected a significantly deeper transit in the
sodium band relative to the adjacent bands.  
We interpret the signal as due to absorption by sodium
in the planetary atmosphere.  We are encouraged
in this interpretation by the following two considerations:
First, a sodium feature in the transmission spectrum of HD~209458
was predicted unanimously by the current
modeling efforts \citep{sea00, bro01a, hub01}.  Second,
observed spectra of L-type objects \citep[e.g.][]{kir99}, 
T-dwarfs \citep[e.g.][]{burg99} and, in particular, the brown dwarf
Gl~229~B \citep[e.g.][]{opp98}, show very
strong absorption from alkali metal lines \citep[][]{bur00}.  
These objects span a temperature range of 900-2000~K, 
which includes the equilibrium temperature of 
HD~209458~b, $T_{eff} = 1430~(1-A)^{\frac{1}{4}}$~K
(where $A$ indicates the Bond albedo).
(It should be noted that these objects are at roughly the same
temperature as HD~209458~b for very different reasons:  
The planet is heated by its proximity to the parent star,
whereas the temperatures for the other objects
are set by their ongoing contraction.)

In the following discussion, we first consider alternate 
interpretations of our result, then discuss the constraints
that we can place on the planetary atmosphere, and finish 
by looking ahead to near-future complementary observations.

\subsection{Alternate Explanations}
We must take care that the observed decrement is not
the result of the distinctive limb-darkening
exhibited by the sodium line relative to that
of the adjacent continuum.  In order to quantify
the amplitude of this effect, we generated model
transit curves (for a planet of \emph{constant} radius)
as described in \S~2.1, and integrated
these in $\lambda$ over the nine band passes listed
in Table~1.  We then recreated the linear combination
of these results as given in eq~(4).  We found that the
predicted offsets were:
\begin{equation}
\begin{array}{cccc}
\Delta n_{Na} & {\rm (theory; constant}~R_{p}{\rm )} & = & +1.52 \times 10^{-5}\phm{.}\\
\Delta m_{Na} & {\rm (theory; constant}~R_{p}{\rm )} & = & +0.39 \times 10^{-5}\phm{.}\\
\Delta w_{Na} & {\rm (theory; constant}~R_{p}{\rm )} & = & +0.47 \times 10^{-5}.
\end{array}
\end{equation}
These values
are much smaller than the signal we detect.
Moreover, they are of the opposite sign: Since the
net decrement toward the limb is less in the
sodium line than in the adjacent continuum, the
transit as observed in a band centered on the sodium
lines would be expected to be \emph{less} deep near $T_{c}$ 
than that observed in a band in the adjacent spectral region.
We note that the above quantities should be subtracted
from the observed values of $\Delta n_{Na}$, $\Delta m_{Na}$, and
$\Delta w_{Na}$ given in equation~(5) for comparison with 
future modeling efforts that do not explicitly include the
wavelength-dependent limb-darkening.  For the purposes
of the present paper, our models explicitly account 
for this effect by making use
of the model stellar surface brightness, and thus
no adjustment is needed.

A second concern is that the star
might appear \emph{smaller} when observed at the wavelength
of the sodium lines, producing a deeper transit for
the same-sized planet.  In
fact, we expect the opposite to be true:
The solar limb should be slightly \emph{larger} in the core of the
Na~D lines because of the great opacity in these lines.  
Model calculations (E.~H.~Avrett, personal communication) yield an
intensity-weighted solar radius increase, after averaging over even
the narrowest bandpass containing the Na~D lines, of less than 10 km.  
This will have a completely negligible effect on the observed transit
depth in the  bandpass containing the Na~D lines.

\subsection{Constraints on the Planetary Atmosphere}
Our fiducial model (s1), which has cloud tops at 0.04~bar and a 
solar abundance of sodium in atomic form, predicts values 
for $\Delta n_{Na}$ and $\Delta m_{Na}$ that are 
$\sim$3 times deeper than we observe.  This conclusion awaits 
confirmation from other more detailed models, such as those performed
by \citet{sea00} and \citet{hub01}.  Here we consider several 
possible physical effects that may contribute to the reduced
sodium absorption relative to the predictions of our model:

It may be that a significant fraction of atomic sodium
has combined into molecules, which may either
be present as gaseous species, or sequestered
from the atmosphere as condensates.
Chemical equilibrium models for substellar and planetary
atmospheres \citep{feg94, feg96, bur99, lod99}
indicate that Na$_{2}$S, NaCl, NaOH, NaH, and more complicated
species (such as NaAlSi$_{3}$O$_{8}$) may be present
in the planetary atmosphere.
In particular, \citet{lod99} finds that, at these temperatures, 
the competition for sodium is predominantly 
between Na$_{2}$S (\emph{solid}) and monatomic Na (\emph{gas}).  
At temperatures above the Na$_{2}$S-condensation line, 
sodium is predominantly in atomic form,
and gases such as NaCl, NaOH, and NaH never become the
most abundant sodium-bearing species (K.~Lodders, personal communcation).   
If the disparity between the observed values for $\Delta n_{Na}$
and $\Delta m_{Na}$ and the predictions for these quantities
based on our fiducial model (s1), 
is due entirely to this effect (and not, for example, mitigated
by the effect of high cloud decks), then this would
indicate that less than 1~\% of the sodium is present
in atomic form.  This statement is based on the
fact that even our model c3 fails to reproduce the observed
values.

Our current models \citep{bro01a} account for the ionization
of sodium by the large stellar flux incident upon the
planetary atmosphere.  However, it is possible
that the effect is larger than predicted by our calculations:
This would require a sink for free electrons, the 
effect of which would be to lower the recombination rate.

Perhaps HD~209458 began with a depleted sodium abundance, 
or, more generally, a depleted metal abundance.  The current
picture of gas-giant planet formation 
predicts that planets should present a metallicity
at or above the value observed in the parent star;
Jupiter and Saturn are certainly metal-rich relative to
the Sun \citep[see discussion in][]{feg94}.
Metallicity appears to play an important role
in the formation of planets: 
Parent stars of close-in planets appear to be metal 
rich \citep{gon97, gon00}
relative to their counterparts in the local solar
neighborhood \citep{que00b, but00}.  Furthermore, 
the observed lack of close-in giant planets in the globular 
cluster 47~Tuc \citep{gil00} may be due, in part, 
to a reduced planet-formation rate due to the 
low-metallicity of the globular cluster.  
The relationship linking 
metallicity and planet formation is not clear.

Clouds provide a natural means of reducing the sodium
absorption, by creating a hard edge to the atmosphere,
and thereby reducing the effective area of the atmosphere
viewed in transmission.  If the difference
between the observed values of $\Delta n_{Na}$
and $\Delta m_{Na}$ and the predictions for these quantities
based on our fiducial model (s1) is due entirely to clouds, 
then this would require a very high cloud deck, with cloud
tops above 0.4~mbar (model n4).  \citet{ack01} have recently
produced algorithms to model cloud formation in the 
atmospheres of extrasolar planets.
The effect of the stellar UV flux on the atmospheric
chemistry has not been explicitly included.  We urge investigations 
of the production of condensates, such as photochemical hazes, 
by this mechanism.  Such 
photochemical hazes are present in the atmospheres of
the gas giants of the 
solar system, and produce large effects upon the 
observed reflectance spectra of these
bodies \citep{kar94}.  The stellar flux incident
upon the atmosphere of HD~209458~b is roughly 20,000 
times greater than that upon Jupiter.

\subsection{Future Observations}
In the present work, we have demonstrated that by using existing
instruments, it is possible to begin the investigation
of the atmospheres of planets orbiting other
stars.  Wide-field surveys conducted by dedicated,
small telescopes \citep[e.g.][]{bor01, bro00}\footnote{See also
http://www.hao.ucar.edu/public/research/stare/stare.html, maintained
by T.~Brown and D.~Kolinski.} promise to detect numerous close-in,
transiting planets over the next few years.  The
parent stars of these planets should be sufficiently
bright to permit similar studies of the planetary atmosphere.

Based on the results presented here, we believe the observed
value of sodium absorption from the planetary atmosphere
may have resulted from a high cloud deck, a low atomic sodium 
abundance, or a combination of both.
Observations of the transmission spectrum over 
a broader wavelength range should allow us to distinguish between these
two broad categories of models (Figure~8).  If the effect is
predominantly due to clouds, the
transmission spectrum from 300--1000~nm will 
show little variation with wavelength,
other than additional alkali metal lines.  If, on the other hand,
the clouds are very deep or not present, then the transmission
spectrum may show deep features redward of 500~nm due to water,
and absorption due to Rayleigh scattering blueward of 500~nm.  These
are only two extremes of the many possibilities; the actual
case may share characteristics of both of these effects.
Further diagnostics could be provided
by near-IR spectroscopy, which could detect features due to molecular
species (such as ${\rm H_{2}O}$, 
CO, and ${\rm CH_{4}}$) that are expected to be present in
planetary transmission spectra \citep{sea00, bro01a, hub01}.
Suitable observations in the 1--5~$\mu$m region would allow inferences
about the atmosphere's thermal structure and composition, permitting
a more definite interpretation of the sodium signature.
\citet{bro00aas} have presented preliminary near-IR spectra 
of HD~209458 in and out of transit.  These spectra demonstrate
that it should be possible to search for near-IR atmospheric signatures
with existing instruments, such as the NIRSPEC spectrograph \citep{mcl98} 
on the Keck~II telescope.  

Observations of the reflected light spectrum (wavelength-dependent
geometric albedo) as pursued for the $\tau$~Boo system 
\citep{cha99, cam99} would complement the results
of transmission spectroscopy.  Predictions for the albedo
as a function of wavelength vary by several orders of magnitude
\citep{sea98, mar99, gou00, sud00}.  These observations should
be facilitated in the case of a transiting system \citep{cn00}.
If possible, the observation of the phase function (variation
in the reflected-light with orbital phase) would constrain
greatly the nature of particulates in the planetary atmosphere 
\citep{sea00a}.  Observations of molecular absorption features
from thermal emission spectra, such as sought in the recent study of the
$\tau$~Boo system \citep{wie01}, would clearly complement
these studies.  Finally, infrared photometry of the secondary
eclipse would provide a direct measurement of the effective
temperature of the planetary atmosphere, should the requisite
precision be achieved.

\acknowledgements
We thank Steven Beckwith for the grant of
Director's Discretionary Time for this project. 
We are grateful to the \emph{HST} STIS and operations teams, especially 
Helen Hart, Gerard Kriss, and Jeff Valenti for their prompt and 
insightful help in resolving the database error experienced during 
the first transit.
We thank the referee William Hubbard for his comments,
as well as for his prompt review of the manuscript.  
We also thank the many theorists, including Travis Barman, Adam Burrows,
Katharina Lodders, Mark Marley, Dimitar Sasselov, and Sara Seager,
for their thoughts on the constraints that could be placed
on the planetary atmosphere.
Support for proposals HST-GO-08789.01-A and HST-GO-08789.02-A was 
provided by NASA through a grant from the Space Telescope Science Institute, 
which is operated by the Association of Universities for Research 
in Astronomy, Inc., under NASA contract NAS5-26555.  Further support
for this work was provided through NASA grant NAG5-10854.

\clearpage

\begin{deluxetable}{cccc}
\tablecaption{Wavelength Bands}
\tablewidth{0pt}
\tablehead{
\colhead{width} & \colhead{blue (nm)} & \colhead{center (nm)} & \colhead{red (nm)}}
\startdata
narrow & $n_b$: 581.8--588.7 & $n_c$: 588.7--589.9 & $n_r$: 589.9--596.8\\
medium & $m_b$: 581.8--587.4 & $m_c$: 587.4--591.2 & $m_r$: 591.2--596.8\\
wide & $w_b$: 581.8--584.3 & $w_c$: 584.3--594.3 & $w_r$: 594.3--596.8
\enddata
\end{deluxetable}

\begin{deluxetable}{ccc}
\tablecaption{Model Atmosphere Parameters}
\tablewidth{0pt}
\tablehead{
\colhead{identifier} & \colhead{cloud tops} & \colhead{metal abundance}\\
\colhead{ } & \colhead{(bar)} & \colhead{(relative to solar)}}
\startdata
s1 & 0.0368\phn\phn & 1.00\\ 
n1 & 3.68\phn\phn\phn\phn & 1.00\\
n2 & 0.368\phn\phn\phn & 1.00\\
n3 & 0.00368\phn & 1.00\\
n4 & 0.000368 & 1.00\\
c1 & 0.0368\phn\phn & 2.00\\
c2 & 0.0368\phn\phn & 0.50\\
c3 & 0.0368\phn\phn & 0.01\\
\enddata
\end{deluxetable}

\clearpage

\begin{figure}
\plotone{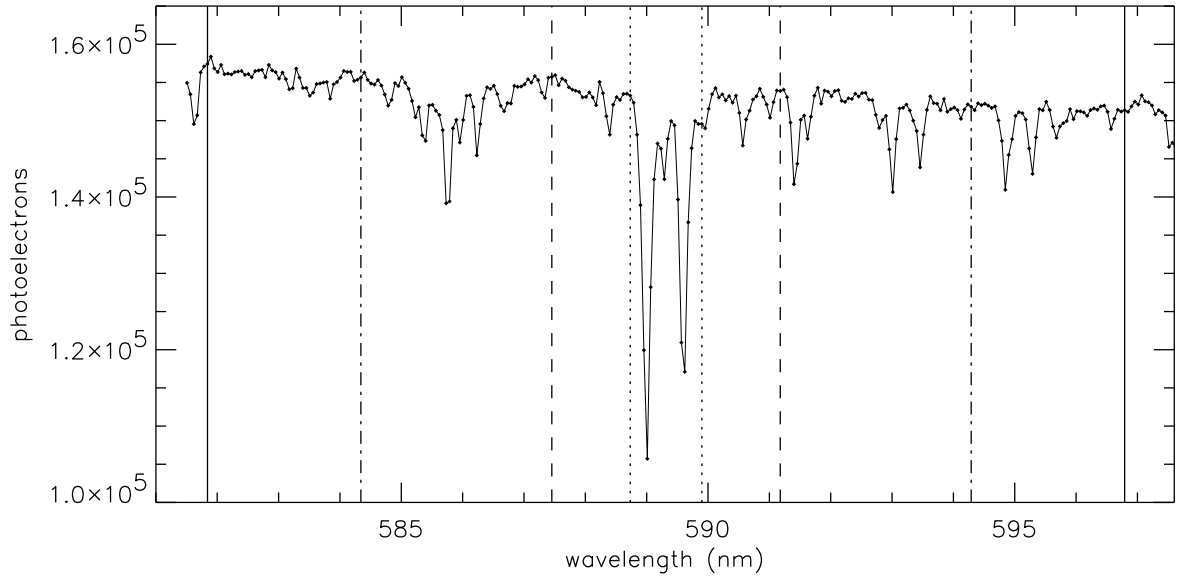}
\figcaption{Shown is a portion of a STIS spectrum of HD~209458,
centered on the Na~D lines.  
The vertical axis is the number of detected photoelectrons per
wavelength pixel after integrating over 17~pixels
in cross-dispersion.  The vertical lines indicate the band 
passes over which we
integrate the spectrum to produce the photometric time series.
The band $n_c$ is the set of pixels between the two dotted
lines; $m_c$ is the set between the dashed lines;
$w_c$ is the set between the dot-dashed lines.  The 
corresponding blue bands ($n_b$, $m_b$, and $w_b$) are the set of
pixels between the left solid line and the left
boundary of the center band.  Similarly, the corresponding
red bands ($n_r$, $m_r$, and $w_r$) are the set of pixels
between the right edge of the center band and the right solid line.}
\end{figure}
\clearpage

\begin{figure}
\plotone{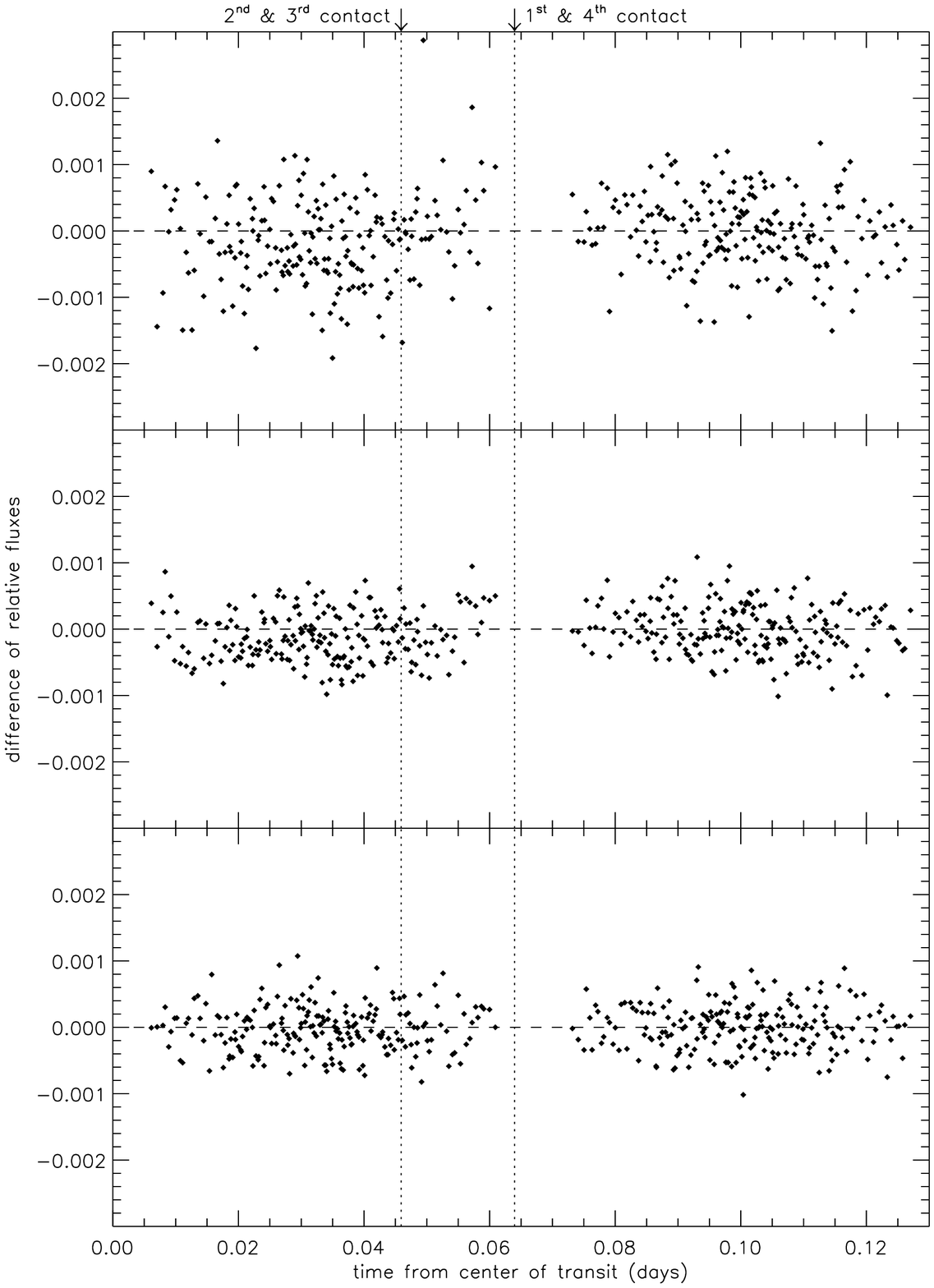}
\end{figure}
\clearpage
\figcaption{(Previous page)~~The unbinned time series $n_{Na}$, $m_{Na}$,
and $w_{Na}$ are plotted in the top, center, and bottom panels,
respectively, as a function of absolute time from the center
of transit.  The mean of the in-transit values of $n_{Na}$
and $m_{Na}$ are both significantly offset below 0.}
\clearpage

\begin{figure}
\plotone{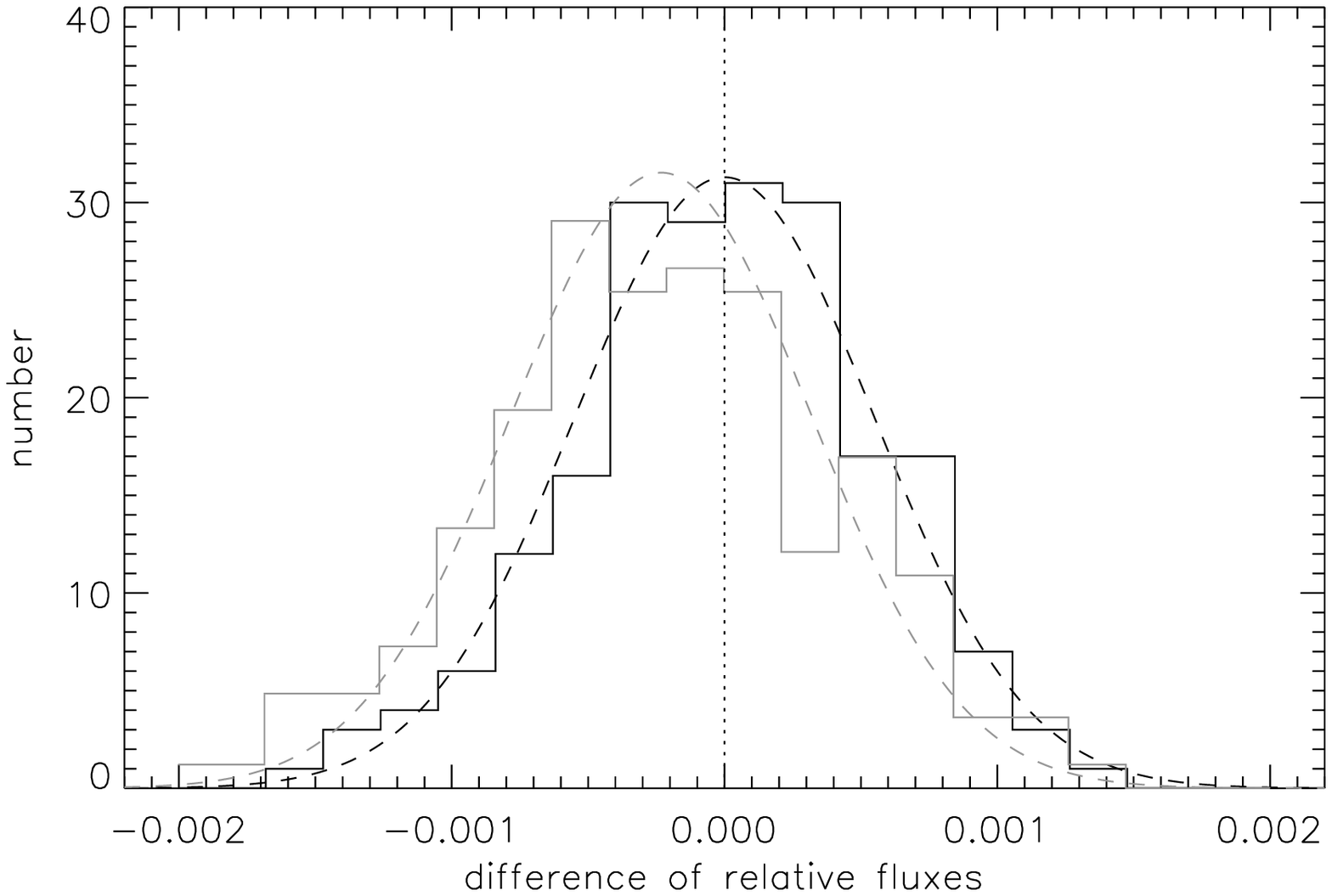}
\figcaption{The dark solid line is a histogram of the out-of-transit
values of $n_{Na}$ (the data shown in the upper panel of Figure~2).  
The dashed dark curve is a Gaussian distribution
with a mean of 0, $\sigma = 5.5 \times 10^{-4}$ as prescribed 
from photon-noise-predictions, and normalized to the same area.  
The gray solid line is a
histogram of the in-transit values of $n_{Na}$, renormalized to the same
number of observations as the dark solid line.  The dashed gray curve
is a Gaussian distribution with the same $\sigma$ as the dark curve, 
but with a mean of \mbox{$-2.32 \times 10^{-4}$} (the observed value).
The gray histogram is significantly offset from 0.}
\end{figure}
\clearpage

\begin{figure}
\plotone{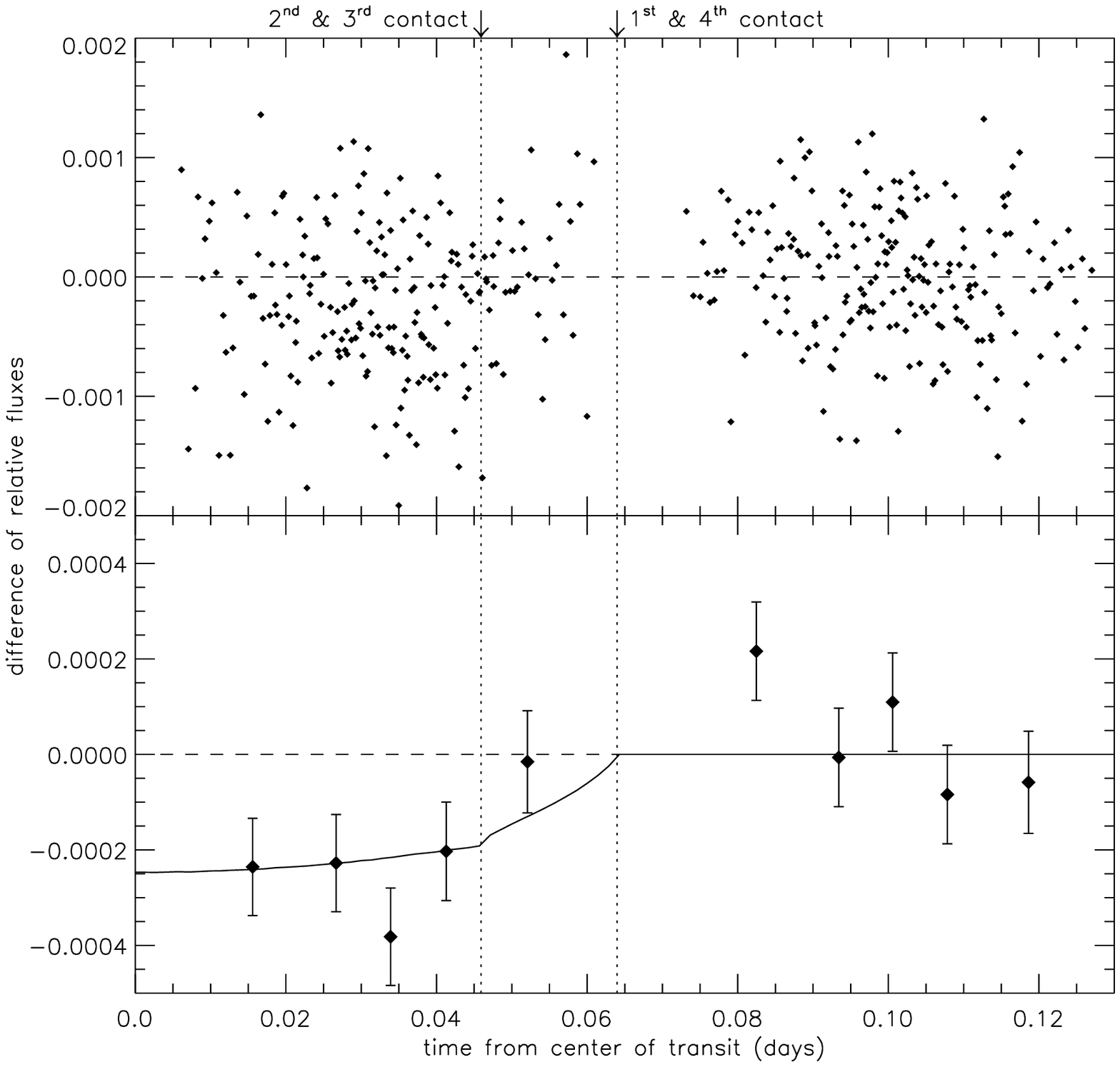}
\figcaption{The upper panel shows the unbinned time series $n_{Na}$
(the data shown in the upper panel of Figure~2).  The lower panel shows these 
data binned in time (each point is the median value in each bin).
There are 10~bins, with roughly equal numbers of observations per bin (42).
The error bars indicate the estimated standard deviation of the median.  
The solid curve is a model for the difference of two transit
curves (described in \S~3), scaled to the observed offset in the mean
during transit, $\Delta n_{Na} = -2.32 \times 10^{-4}$.}
\end{figure}
\clearpage

\begin{figure}
\plotone{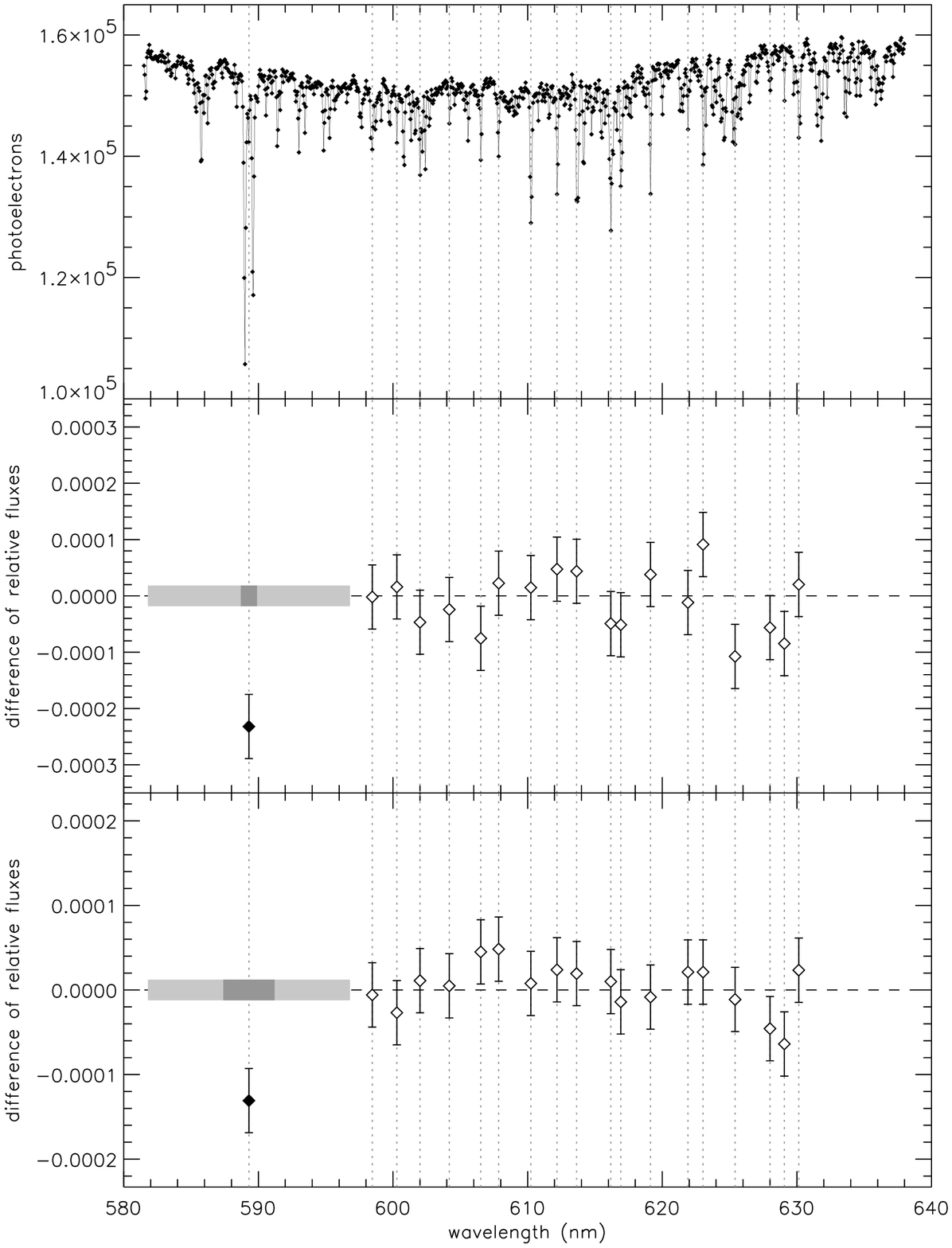}
\end{figure}
\clearpage
\figcaption{(Previous page)~~The upper panel shows a typical 
spectrum of HD 209458, over the full spectroscopic range 
that we observed.  We selected 18 strong features in 
addition to the Na~D
lines; these are indicated by vertical dashed lines.
The middle panel shows the results for the narrow ($n$)
band pass.  The black diamond is the value of $\Delta n_{Na}$.  
The wavelength ranges of the bands $n_{b}$, $n_{c}$, and $n_{r}$
are shown as the horizontal gray bars.  We then repeated
the analysis, i.e. same band widths, but now centered
over each of the 18 features that are indicated.  The resulting
values are shown as white diamonds.
The lower panel shows the corresponding results for
the middle ($m$) band pass.  
In each case, the white diamond points
appear to be normally-distributed, with a mean of 0, and a standard
deviation as prescribed by photon statistics.  In particular, 
the residuals do not correlate with the depth of the spectral feature. 
Some correlations between adjacent white points are
evident, as expected, since the band passes
are wider than the separation between points.
In the middle and lower panels, the black diamond point 
is inconsistent with no variation at the level of 
4.1~$\sigma$ and 3.4~$\sigma$, respectively.  Note that
the y-axis in the lower panel has been scaled so that
the error bars appear the same size as in the middle
panel.}
\clearpage

\begin{figure}
\plotone{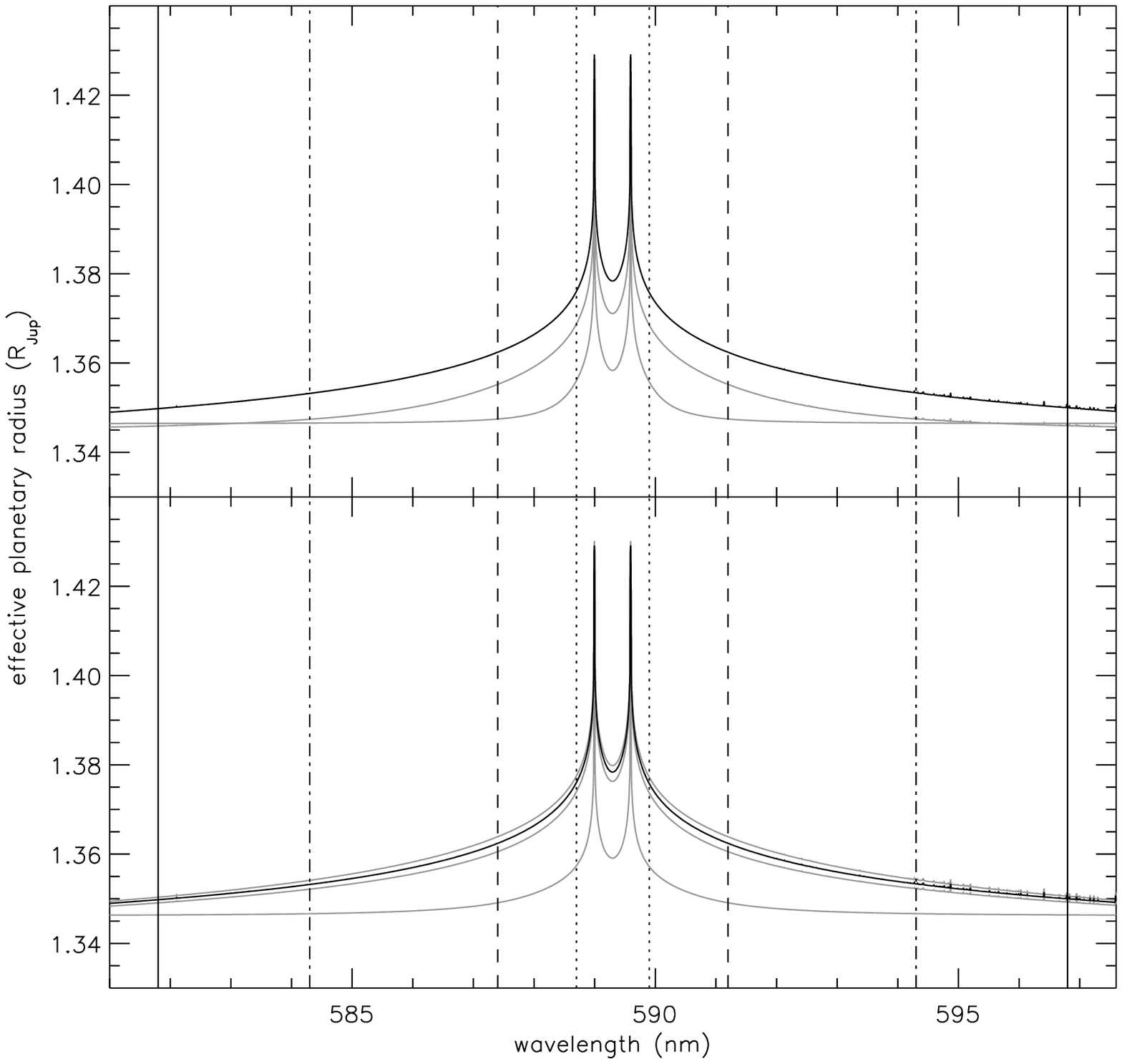}
\figcaption{The effective planetary radius is shown
as a function of wavelength for a variety of atmospheric
models; the vertical lines indicate the bandwidths,
and are the same lines as those in Figure~1.
In both panels, the dark line shows the planetary
radius for the fiducial model (s1).  In the upper panel,
we show the variation with cloud height; the upper and
lower gray curves are models n3 and n4, respectively. 
(The results for models n1 and n2 are indistinguishable 
on this plot from that for s1). In the lower panel, 
we show the variation with sodium abundance; the gray
curves are, from top to bottom, models c1, c2, and c3.}
\end{figure}
\clearpage

\begin{figure}
\plotone{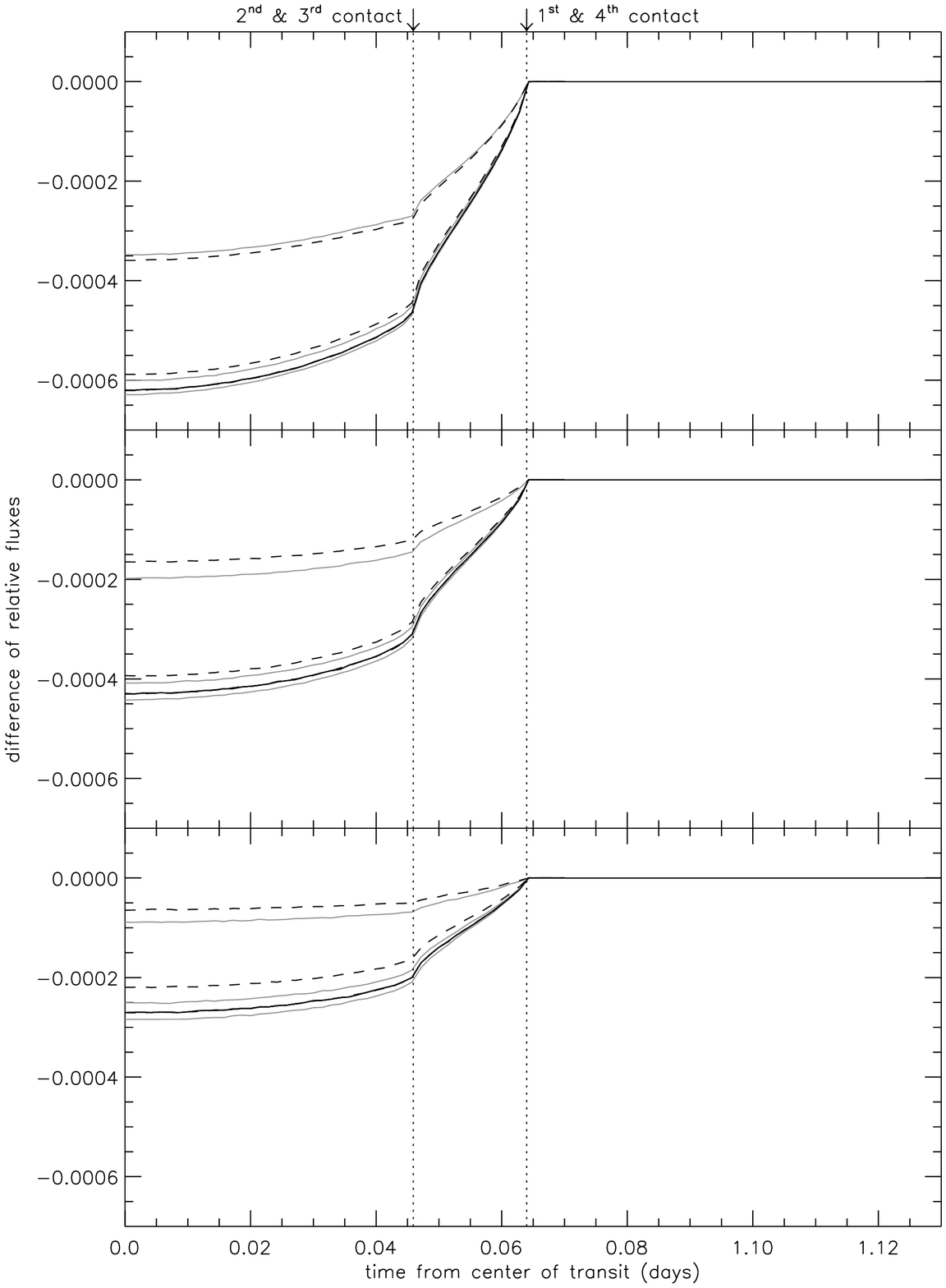}
\end{figure}
\clearpage
\figcaption{(Previous page)~~The solid curve in the top panel
shows the model curve for $n_{Na}$ for the
fiducial model (s1).  The solid curves in the
middle and lower panels are the results for
$m_{Na}$ and $w_{Na}$, respectively, for this
same model.  The dashed lines 
show the variations due to cloud height: 
Within each panel, the upper dashed curve is for model n4, 
and the lower dashed curve is for model n3.  
(The results for models n1 and n2 are 
indistinguishable on this plot from that for s1).  
The solid gray lines show the variations
due to atomic sodium abundance:
Within each panel, the gray curves correspond 
(from top to bottom) to models c3, c2, and c1.  
These curves may be compared directly with
the data presented in Figures~2 and 4.}
\clearpage

\begin{figure}
\plotone{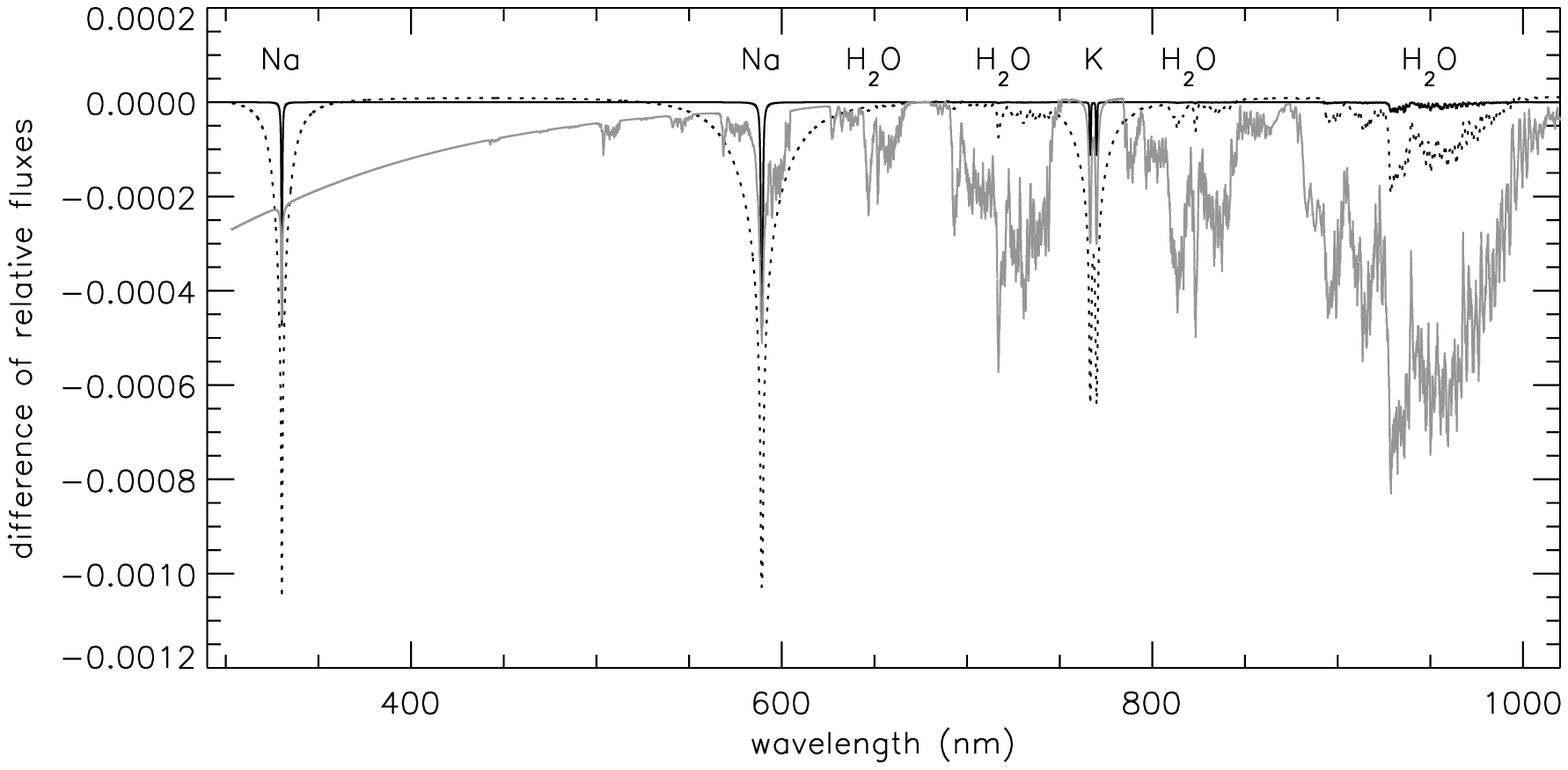}
\figcaption{Shown are three model transmission spectra for HD~209458
over a wavelength range that exceeds that of our 
current data set.
The fiducial model (s1) with a solar abundance of sodium, and cloud
tops at 0.0368~bar, is shown as a dotted line.  This model is excluded
by the data, which do not permit such a deep sodium feature.  
One possibility is that very high clouds reduce the depth of the
sodium absorption feature; such a model is shown as the solid line.  
Another possibility is that the abundance of atomic sodium in the planetary 
atmosphere is reduced greatly from the solar value; this is shown as 
the gray line.  Observations of the transmission spectrum over the
range shown here should allow for us to distinguish between these
two broad categories of models.}
\end{figure}

\end{document}